\documentclass[twocolumn]{article}

\pdfoutput=1
\usepackage{pgf, pgfplots, adjustbox}
\usepackage{amsmath, amssymb, tikz}
\usepackage{float}
\usepackage{makecell}

\usepackage{hyperref}
\usepackage[margin=2cm]{geometry}

\def\<{\langle}
\def\>{\rangle}

\begin{document}
\title{Learning to detect entanglement}
\date{March 20, 2024}
\author{Bingjie Wang\footnote{Email: {\tt bingjie [dot] wang [at] columbia [dot] edu}}}
\pagenumbering{gobble}

\maketitle

\begin{abstract}
Classifying states as entangled or separable is a fundamental, but expensive
task. This paper presents a method, the forest algorithm, to improve the
amount of resources needed to detect entanglement. Starting from ``optimized''
methods for using geometric criterion to detect entanglement, specific steps 
are replaced with machine learning models. Tests using numerical simulations 
indicate that the model is able to declare a state as entangled in fewer steps 
compared to existing methods. This improvement is achieved without affecting
the correctness of the original algorithm.
\end{abstract}

\section{Introduction}
Entanglement is a defining feature of quantum information. Detection algorithms
are typically based on either quantum state tomography or entanglement witnesses
\cite{es1, es2}. Witnesses can reveal entanglement in a single measurement,
but only for a small portion of states. This drawback makes witnesses
difficult to use without prior information about the underlying state. Tomography
does not require prior information, but the number of measurements quickly
becomes unreasonable as the number of qubits increases.

In this paper, pure states are denoted by $|\varphi\rangle$ and mixed states by
$\rho$. A separable pure state admits the factorization:
$|\varphi\rangle = |\phi_1\> \otimes |\phi_2\> \otimes \cdots \otimes 
 |\phi_N\>$, where $N$ is the number of qubits. Similarly, separable mixed 
states can be expressed in the following form:
\begin{equation}
\label{factorizable}
  \rho = \sum_i p_i \rho_1 \otimes \rho_2 \otimes \cdots \otimes \rho_N
\end{equation}
where $\sum_i p_i = 1$. Non-separable states are entangled.

Quantum state tomography is the following problem: given a set of 
measurement results on sets of positive operator value measures, 
infer the state $\rho$. Given a source producing an unknown quantum state, to 
determine whether the state is entangled, one could observe the source, infer the
state's description via tomography, and compute whether the state can be written 
in the form of Eq. \eqref{factorizable}.

Consider the space of linear operators acting on a $D$-dimensional Hilbert space. 
This space has an orthonormal basis, $\{B_i\}$, for the inner product 
$\<B_1, B_2\> = \mathsf{tr}(B_1^\dagger B_2)$. Under this basis, the 
state $\rho$ can be represented as:
\begin{equation}
\label{param_est}
   \rho = \frac{1}{D}\cdot \mathsf{id} + \sum_{i = 1}^{D^2 - 1} x_i B_i
\end{equation}
where $\mathsf{id}$ is the identity operator and $B_0$ is taken to be 
$\mathsf{id}/\sqrt{D}$. Here, $x_i = \mathsf{tr} (\rho^\dagger B_i)$, is the 
expected value of the measurement associated with $B_i$. This paper uses 
$\<B_i\>$, with the state $\rho$ implicit, to represent this expectation.

For $N$-qubit quantum states, it is customary to use tensor products of Pauli
matrices, $\{\mathsf{id}, \sigma_1=\sigma_x, \sigma_2=\sigma_y, \sigma_3=\sigma_z\}$, as the basis. According to the geometric criterion
\cite{geometric}, a state is entangled if:
\begin{equation}
\label{geom_eq}
    \sum_{a_1 = 1}^3 \sum_{a_2 = 1}^3 \cdots \sum_{a_N = 1}^3
    \langle \sigma_{a_1} \otimes \sigma_{a_2} \otimes \cdots \otimes
            \sigma_{a_N} \rangle^2 > 1
\end{equation}
For a state $\rho$, write $\overline{x}(\rho)$ as the vector of expectations
corresponding to measurements associated with elements of the {\it improper basis}
consisting of tensor products of Pauli matrices excluding the identity. For
vectors of reals, $\overline{s} \cdot \overline{e} < \overline{e} \cdot
\overline{e}$ implies $\overline{s} \neq \overline{e}$. Thus, a state cannot be
equal to a separable state if:
\begin{equation}
\label{geom_intuit}
\overline{x}(\rho) \cdot \overline{x}(\rho) > 
\max_{\text{separable} |\varphi\>} 
	\overline{x}(\rho) \cdot \overline{x}(|\varphi\>\<\varphi|)
\end{equation}
It can be shown that the right hand side of Eq. \eqref{geom_intuit} is less or equal to 1.

There are $2^{2N} - 1$ elements in the Pauli basis and as many parameters for 
quantum state tomography to infer. This is prohibitively expensive.
With the geometric criterion, only a (often much smaller) subset, such
that the inequality in Eq (\ref{geom_eq}) is violated, is required. Following the
convention in \cite{tree}, an measurement is ``large'' or ``small'' based on the
the contribution it makes to violating the geometric criterion.

Laskowski {\it et al} proposed a method that eliminates measurements that are 
likely to be small by exploiting correlation complimentarity \cite{tree, complimentarity}, 
which states, for a mutually anti-commuting set of operators,  $\mathsf{S}$:
\begin{equation}
\label{complimentarity}
    \sum_{B \in \mathsf{S}} \langle B \rangle^2 \leq 1
\end{equation}
This method, the tree algorithm, uses a search tree to navigate mutually 
anti-commuting subsets of the improper basis. If an inferred $\<B\>^2$ is large,
({\it i.e.}, $>0.16$) the tree moves onto a different subset.

The main idea of this paper is to replace the search tree with a random forest, 
a type of machine learning model. The training process to build
the model and how the model can be used to select which expectation to
infer next is described. The model does not make decisions about
whether a state is entangled and thus any state that violates the geometric criterion
will be declared as entangled. Numerical evidence is presented 
to indicate that the model is more likely to select more helpful measurements.

\section{Forest Algorithm}
\label{forest}
In (supervised) machine learning, the input can be described by a vector
$\overline{x}$ and the output by $y$. Given a training set of $(\overline{x}, y)$
samples, the machine learning algorithm learns the input-output relationship,
$f(\overline{x})$, by fitting $f$ to the training set. A more comprehensive and
detailed treatment is presented in \cite{islr}.

When using the geometric criterion, the entanglement detection procedure is:
\begin{enumerate}
\itemsep0em
	\item There is a source producing a quantum state. Initially, nothing is known
about the state.
	\item Choose $B$ from the improper basis, using a score based on the expected 
value of $\<B\>^2$.
	\item Repeatedly make the measurement associated with $B$ to determine
$\<B\>$. The knowledge obtained from these measurements are used to update the
scores.
	\item Repeat steps 2 and 3 until the sum of inferred $\<B\>^2$s exceeds 1.
\end{enumerate}

As there is no initial information about the state, the natural choice of how to obtain
the training set is by sampling pure states using the Haar measure.

The challenge of expressing step 2 in the machine learning framework is that the
amount of information to available predict a given $\<B\>^2$ changes. For now, 
consider the ``last iteration''. As an example, using $\<ab\>$ as a shorthand for
$\sigma_a \otimes \sigma_b$, the model could be asked to make a judgement about 
$\<xx\>^2$. At the last iteration, all the information that could be used to predict
$\<xx\>^2$, namely, $\{\<xy\>^2, \<xz\>^2, \cdots, \<zz\>^2\}$, would be 
known (presuming the geometric criterion has not yet been satisfied). However, any
element of the improper basis could be the last to be selected. Thus, for two qubits,
9 models need to be trained:
\begin{equation}
\begin{cases}
	f_{xx}: \overline{x}_{xx} &= [\<xy\>^2, \<xz\>^2, \cdots, \<zz\>^2], \\
	\phantom{f_{xx}:\,\,} y_{xx} &= \mathbb{I}(\<xx\>^2 \geq \max_i(\overline{x}_{xx})_i) \\
	f_{xy}: \overline{x}_{xy} &= [\<xx\>^2, \<xz\>^2, \cdots, \<zz\>^2], \\
	\phantom{f_{xy}:\,\,}y_{xy} &= \mathbb{I}(\<xy\>^2 \geq \max_i(\overline{x}_{xy})_i) \\
	& \cdots  \\
	f_{zz}: \overline{x}_{zz} &= [\<xx\>^2, \<xy\>^2, \cdots, \<zy\>^2], \\
	\phantom{f_{zz}:\,\,}	y_{zz} &= \mathbb{I}(\<zz\>^2 \geq \max_i(\overline{x}_{zz})_i) 
\end{cases}
\end{equation}
where $\mathbb{I}$ is the indicator function. Generalizing: 
\begin{equation}
\label{proxy}
f_j \sim \mathbb{I}\left[ x_j = \max_i
	\overline{x}(\rho)_i \text{ given } \overline{x}(\rho)_{-j} \right]
\end{equation}
where $\overline{x}_{-j}$ is the vector $\overline{x}$ without entry $j$. The notation
$f \sim g$ indicates $f$ is trained to learn $g$. 

The choice of $y$ as the maximum reflects the fact that the algorithm only considers
which choice is best, not how much better the best choice is compared to the second-best
choice.

It remains to choose a functional form for $f$ and fit $f$ to the data. A random forest,
which is a collection of decision trees, is used here. Formally, a decision tree 
partitions the input space into regions, $r_1, r_2, \cdots r_k$, with associated 
output values, $y_1, y_2, \cdots, y_k$, and
\begin{equation}
  f^{\text{tree}}(\overline{x}) = \sum_{i = 1}^k y_i \cdot \mathbb{I}(\overline{x} \in r_i)
\end{equation}
The forest's output is the average (or majority vote) of its constituent decision trees.  

The training procedure is discussed extensively in \cite{islr}. A sketch
of the construction for binary classification problems, where $y \in \{0, 1\}$,
is as provided. To train a tree, the tree initially has one region containing the entire 
input space.  Samples falling into a region is viewed as a data 
producing $1$s and $0$s and entropy can be calculated for this source. 

During each iteration, each region is split by a decision rule $x_j < \theta$.
Let $p(r)$ and $n(r)$ be the number of $y = 1$ and $y=0$ samples in the training 
set falling, respectively, into region $r$, and write $t(r)$ for $p(r) + n(r)$. 
Then let $L(p(r), n(r))$ be an impurity measure (entropy is used, but the Gini Index 
is also popular). Now let $r$ be the original region and $r_1$ and $r_2$ be the 
regions after the split, define the (ascending) impurity gain for a split as:
\begin{equation}
\label{impurity}
\begin{aligned}
	G(r, j, \theta) = &t(r)L(r) - \\ 
	&\left[t(r_1)L(r_1) + t(r_2)L(r_2)\right]
\end{aligned}
\end{equation}
Since the goal is for the tree to become more pure when descending the tree, the 
decision rule is chosen to maximize the impurity gain going downwards.

Iterative splitting continues until all regions are (mostly) pure. To build a forest, train 
many trees on carefully selected subsetsof the training data such that the predictions 
of each tree are anti-correlated. This stabilizes predictions by reducing the variance 
on the average.

With the trained model in hand, it is time to tackle the problem of updating the scores
based on the measurement results. Let $R_j$ be the set of input vectors for model
$f_j$ that agree with the known information. Define the tree's score as:
\begin{equation}
\label{tree_score}
	s^{\text{tree}}_j = \frac{
		\sum_{r_i \cap R_j \neq \emptyset} p(r_i)
	}{\sum_{r_i \cap R_j \neq \emptyset} t(r_i)} 
\end{equation}
The forest's score is the average of its constituent decision trees' scores. To explain 
the notation $r_i \cap R_j \neq \emptyset$, suppose the only known measurement
so far is $\<zz\>^2 = 0.25$. Now consider a region is defined by the decision rule 
$\<zz\>^2 \geq 0.16$, then the intersection of vectors with $\<zz\>^2 \geq 0.16$ and 
vectors with $\<zz\>^2 = 0.25$ is all quantum states with $\<zz\>^2 = 0.25$,
a non-empty set. However, the other subdivided region with $\<zz\>^2 < 0.16$ does
have an empty intersection. When the denominator of the Eq (\ref{tree_score}) is 0,
then $s^{\text{tree}}_j$ is defined as 1. 

\begin{figure}
\label{toy_tree}
\begin{tikzpicture}
  \node[shape=circle, draw=black] (yy) at (0,0) {$yy$?};
  \node[shape=circle, draw=black] (zz) at (1.75, -1.75) {$zz$?};
  \node[shape=circle, draw=black] (p) at (3.5, -3.5) {$6, 3$};
  \node[shape=circle, draw=black] (n) at (0, -3.5) {$3, 2$};
  \node[shape=circle, draw=black] (p1) at (-1.75, -1.75) {$1, 5$};
\begin{scope}[every node/.style={fill=white}, 
       every edge/.style={thick, draw=black}]
  \path[->] (yy) edge node {$\geq0.25$} (zz);
  \path[->] (yy) edge node {$<0.25$} (p1);
  \path[->] (zz) edge node {$\geq 0.16$} (p);
  \path[->] (zz) edge node {$<0.16$} (n);
\end{scope}
\end{tikzpicture}
\caption{
\label{toy_tree}
  A ``toy'' tree. A decision rule $\<ab\>^2 < \theta$ is represented as a node $ab?$ 
  with $<$ and $\geq$ edges for each possibility. A terminal node represents a region
  defined $r$ defined by the decision rules on the path leading to the node and the 
  node is labelled with the pair $p(r), n(r)$. 
}
\end{figure}

For a more complete example, consider a ``toy'' tree for $f_{xx}$, represented as a graph in 
Fig. \ref{toy_tree}. Suppose, so far, the results are $\<yy\>^2 = 0.36$ and 
$\<zz\>^2 = 0.01$. Then, the tree's score is: $3 / (3 + 2) = 0.6$ as only the centre 
node is reachable.

\section{Numerical Experiment}
The performance of the forest algorithm is analyzed by generating two to six qubit states
and running both the forest and tree algorithms on the same states. Since the tree
algorithm requires $\<xx\cdots x\>$ to be large, a rotation is applied to enforce this
criterion. Also included in the 
comparison are: ``optimal'' and ``random''. Optimal always makes the 
largest unknown measurement, whereas random chooses an unknown measurement at 
random, but starting with $\<xx\cdots x\>$. 

Source code for the numerical experiments is available at \cite{code}. The forests were 
generated using
the {\tt scikit-learn} package \cite{scikit}  and {\tt Eigen} package \cite{eigen} was 
used for matrix computations. Due to practical limitations, 30000 states were used
to train each model and 1500 states were tested for each qubit.
Machine learning models generally ``get better'' with more training data and having 
more data could improve the result.

The test sets of $N$-qubit states were constructed using the following procedure:
\begin{equation}
|\varphi\> = R \cdot G_1G_2\cdots G_n \left(|0\>^{\otimes N}\right)
\end{equation}
where $n$ is a random integer between $K = 3\cdot N^2 + 27$ and $2 \cdot K$ 
and each $G_i$ is a quantum gate operating on 1 or 2 chosen qubits from the 
universal gate set \cite{universal}:
\begin{equation}
\left\{
\frac{1}{\sqrt{2}} \begin{bmatrix} 1 & 1 \\ 1 & -1 \end{bmatrix},\,\,
\begin{bmatrix} 1 & 0 \\ 0 & e^{i\pi/4} \end{bmatrix},\,\,
\begin{bmatrix} 
	1 & 0  & 0 & 0 \\ 
	0 & 1 & 0 & 0 \\ 
	0 & 0 & 0 & 1 \\
	0 & 0 & 1 & 0
\end{bmatrix}
\right\}
\end{equation}
The final rotation, $R$, ensures $\<xx\cdots x\>$ is the largest measurement.
Finally, rejection sampling is applied to discard states that cannot be declared
entangled by the geometric criterion, as all of the measurements would need to be
made regardless of algorithm.

This roundabout construction (as opposed to sampling over a uniform measure) 
is to verify how well the forests ``generalize'' to a generic test distribution of states 
that is different to the distribution that it was trained on. The final
rotation making $\<xx\cdots x\>$ the largest measurement is a major concession
to the tree algorithm that allows the mutually anti-commuting subset strategy to be
applied to its fullest extent.
Despite this concession, the average number of measurements taken before the 
geometric criterion is violated is:
\begin{center}
\begin{tabular}{c | c | c | c | c | c}
Qubits & Forest & Tree & $\Delta$ & Optimal & Random \\ \hline
2 & 4.41 & 4.44 & 0.03 & 2.96 & 4.85 \\
3 & 5.91 & 6.06 & 0.16 & 3.01 & 7.52 \\
4 & 10.65 & 10.94 & 0.29 & 3.80 & 13.77 \\
5 & 21.83 & 23.14 & 1.31 & 5.43 & 27.29 \\
6 & 48.91 & 50.80 & 1.89 & 8.65 & 57.21 \\
\end{tabular}
\end{center}
Detailed results for three to six qubits are presented in Fig. \ref{access_performance}.

These results indicate that the forest algorithm is an improvement over the tree
algorithm. The advantage of the forest algorithm over the tree algorithm increases
with the number of qubits, possibly exponentially. Both algorithms are better than 
random, suggesting that the mathematical structure of entanglement is being exploited. 
It is unclear if
anything close to the ``optimal'' performance can be achieved without further
experimental assumptions.

\onecolumn
\begin{figure}[h]
\begin{tabular}{c c}
    {\bf Three Qubits} & {\bf Four Qubits} \\
    \begin{tikzpicture}
      \pgfplotsset{every x tick label/.append style={font=\footnotesize}}
      \pgfplotsset{every y tick label/.append style={font=\footnotesize}}
      \begin{axis} [xlabel = Number of Measurements, ylabel = States Detected, 
                    legend style = {at={(0.98,0.02)},anchor = south east}]
      \addplot [color=gray] table [x = steps, y = random, col sep = comma] {access3.csv};
      \addlegendentry{random}
      \addplot [mark=none] table [x = steps, y = optimal, col sep = comma] {access3.csv};
      \addlegendentry{optimal}
      \addplot [color=blue] table [x = steps, y = forest, col sep = comma] {access3.csv};
      \addlegendentry{forest}
      \addplot [color=orange] table [x = steps, y = tree, col sep = comma] {access3.csv};
      \addlegendentry{tree}
      \end{axis}
    \end{tikzpicture} &
    \begin{tikzpicture}
      \pgfplotsset{every x tick label/.append style={font=\footnotesize}}
      \pgfplotsset{every y tick label/.append style={font=\footnotesize}}
      \begin{axis} [xlabel = Number of Measurements, ylabel = States Detected, 
                    legend style = {at={(0.98,0.02)},anchor = south east}]
      \addplot [color=gray] table [x = steps, y = random, col sep = comma] {access4.csv};
      \addlegendentry{random}
      \addplot [mark=none] table [x = steps, y = optimal, col sep = comma] {access4.csv};
      \addlegendentry{optimal}
      \addplot [color=blue] table [x = steps, y = forest, col sep = comma] {access4.csv};
      \addlegendentry{forest}
      \addplot [color=orange] table [x = steps, y = tree, col sep = comma] {access4.csv};
      \addlegendentry{tree}
      \end{axis}
    \end{tikzpicture} \\
    {\bf Five Qubits} & {\bf Six Qubits} \\
    \begin{tikzpicture}
      \pgfplotsset{every x tick label/.append style={font=\footnotesize}}
      \pgfplotsset{every y tick label/.append style={font=\footnotesize}}
      \begin{axis} [xlabel = Number of Measurements, ylabel = States Detected, 
                    legend style = {at={(0.98,0.02)},anchor = south east}]
      \addplot [color=gray] table [x = steps, y = random, col sep = comma] {access5.csv};
      \addlegendentry{random}
      \addplot [mark=none] table [x = steps, y = optimal, col sep = comma] {access5.csv};
      \addlegendentry{optimal}
      \addplot [color=blue] table [x = steps, y = forest, col sep = comma] {access5.csv};
      \addlegendentry{forest}
      \addplot [color=orange] table [x = steps, y = tree, col sep = comma] {access5.csv};
      \addlegendentry{tree}
      \end{axis}
    \end{tikzpicture} &
    \begin{tikzpicture}
      \pgfplotsset{every x tick label/.append style={font=\footnotesize}}
      \pgfplotsset{every y tick label/.append style={font=\footnotesize}}
      \begin{axis} [xlabel = Number of Measurements, ylabel = States Detected, 
                    legend style = {at={(0.98,0.02)},anchor = south east}]
      \addplot [color=gray] table [x = steps, y = random, col sep = comma] {access6.csv};
      \addlegendentry{random}
      \addplot [mark=none] table [x = steps, y = optimal, col sep = comma] {access6.csv};
      \addlegendentry{optimal}
      \addplot [color=blue] table [x = steps, y = forest, col sep = comma] {access6.csv};
      \addlegendentry{forest}
      \addplot [color=orange] table [x = steps, y = tree, col sep = comma] {access6.csv};
      \addlegendentry{tree}
      \end{axis}
    \end{tikzpicture}
\end{tabular}
\caption{
  The performance of the tree and forest algorithms on sample of 1500 entangled states. 
 ``Optimal'' refers to the number of measurements required to prove entanglement via 
  the geometric criterion if there were an oracle for the largest measurement. 
  ``Random'' refers to making measurements at random.
  The forest algorithm is able to declare states as entangled with fewer measurements 
  than the tree algorithm on average. Its edge grows with the number of qubits.
}
\label{access_performance}
\end{figure}
\twocolumn

\section{Discussions}
This section aims to provide intuition for the forest algorithm. Following the work
of Hayden, Leung, and Winter in \cite{generic}, concentration of measure arguments
indicate random forests are well suited to capture the relationships between the 
measurements. 

It is well known that a continuous measure that is invariant under unitary 
transformations is a uniform measure. For a group, $(\mathsf{G}, \oplus)$, the 
Haar measure, $\mu$, is the measure such that, for any $S \subseteq G$ and 
$g \in \mathsf{G}$, 
$\mu(\mathsf{S}) = \mu(\{g \oplus s\,|\, s\in \mathsf{S} \}) 
                             = \mu(\{ s \oplus g\,|\, s \in \mathsf{S} \})$.
Sampling unitary matrices from its Haar measure is unitary invariant by 
definition. As expected, the prior represents a uniform distribution.
However, L\'{e}vy's Lemma \cite{concentrationofmeasure} states: if $f$ has 
Lipschitz constant $\eta$ (that is, 
$|f(\mathbf{x}) - f(\mathbf{y})| \leq \eta ||\mathbf{x} - \mathbf{y}||$) and 
$X$ is sampled according to the uniform measure on the
$k$-dimensional sphere, then:
\begin{equation}
  \mathsf{pr}(|f(X) - m_f| > \sqrt{\varepsilon}) \leq
  \exp\left[-(k-1)\varepsilon / (2\pi^2 \eta^2)\right]
\end{equation}
where $m_f$ is the median of $f$ under the same measure. 

Consider the expectation as a function. First, convert 
$|\varphi\rangle$ into the real vector, $\boldsymbol{\varphi} =
  [\boldsymbol{\varphi}^r, \boldsymbol{\varphi}^i]^\mathsf{t}$,
where $\boldsymbol{\varphi}^r$ and $\boldsymbol{\varphi}^i$ are the
vectors containing the real and imaginary parts of $|\varphi\rangle$ respectively.
Then, for a Pauli matrix, $B$:
\begin{equation}
\begin{aligned}
f(\boldsymbol{\varphi}) \langle B \rangle &=
\mathsf{tr}\left[|\varphi\rangle\langle \varphi| \cdot B\right]  \\
 &= \sum_{jk} (\varphi^r_j - i\varphi^i_j)(\varphi^r_k + i\varphi^i_k) B_{kj}
\end{aligned}
\end{equation}
L\'{e}vy's Lemma  can be applied to $f$, as $\boldsymbol{\varphi}$
is now uniformly distributed on the $2^{N+1}$ dimensional sphere. The partial
derivatives of $f$ are:
\begin{align}
  \frac{\partial f}{\partial \varphi^r_m} &= 2\varphi^r_m B_{mm} +
    \sum_{j \neq m} \varphi^*_j B_{mj} + \sum_{k \neq m} \varphi_k B_{km} \\
  \frac{\partial f}{\partial \varphi^i_m} &= 2\varphi^i_m B_{mm} +
    i\sum_{j \neq m} \varphi^*_j B_{mj} - i\sum_{k \neq m} \varphi_k B_{km}
\end{align}
Now, bound $(\nabla f)^\dagger (\nabla f)$:
\begin{equation}
\begin{aligned}
   &(\nabla f)^\dagger (\nabla f)  \\
   &=
    \sum_m \left(\frac{\partial f}{\partial \varphi^r_m}\right)^*
           \left(\frac{\partial f}{\partial \varphi^r_m}\right) +
           \left(\frac{\partial f}{\partial \varphi^i_m}\right)^*
           \left(\frac{\partial f}{\partial \varphi^i_m}\right) \\
    &= 2\sum_m |\varphi_m|^2 \cdot |B_{mm}|^2 +
       2\sum_{jk} \varphi^*_j \varphi_k \left(\sum_m B_{km}B_{mj} \right)
\end{aligned}
\end{equation}
The sum has been simplified by noting that the Pauli matrices are
Hermitian. The first term is bounded by $2\sum_m |\varphi_m|^2 = 2$ 
and the second is 
$\mathsf{tr}[|\varphi\rangle\langle \varphi| \cdot B^2] = 1$, as
Pauli matrices square to identity. This bound is an appropriate value for 
$\eta^2$. Finally, Applying L\'{e}vy's Lemma shows that
$\langle B\rangle^2 < \varepsilon$ with high probability, that is, with 
probabillity $1 - o(1)$.

This result indicates that for a given state, most observables are small,
but a subset of observables are very large (in order to violate the 
geometric criterion). For different states, different observables are large. 
This large/small behaviour is easily captured by the decision rules, making 
random forests a good machine learning algorithm for this task.

The impurity gain of each node indicates that the model captures the concept
of correlation complimentarity. Recalling
Eq. (\ref{impurity}), the importance of $\<B_j\>$ for determining the size of, say,
$\<xx\cdots x\>$ can be computed as the fraction of impurity gain on nodes split 
using the $\<B_j\>$ in the forest for $\<xx\cdots x\>$ compared
to the impurity gain of all nodes in the tree. In the two qubit case of determining 
whether $\<xx\>$ is going to be large, the forest assigns
$\approx$ 16\% importance to $\<xy\>, \<xz\>, \< yx\>$ and $\<zx\>$
but only $\approx$ 9\% importance to the other four features. The heavier
emphasis of anti-commuting measurements is true for other models and for
more qubits.

Further evidence that the forests ``learned'' correlation complimentarity
can be seen in example traces.
Let $|D_1^3\> = \left[ |001\> + |010\> + |100\> \right] / \sqrt{3}$ and
$|D_2^3\> = \left[ |011\> + |101\> + |110\> \right] / \sqrt{3}$.
Consider the execution trace of the forest algorithm on the following state with 
$\alpha = 5\pi / 71$, from the experiments in \cite{tree}.
\begin{equation}
  |G(\alpha)\rangle = \cos(\alpha) |D_2^3\rangle + \sin(\alpha)|D_1^3\rangle
\end{equation}

\begin{center}
\begin{tabular}{c | c | c | c }
$i$ & $B_i$ & $\langle B_i \rangle$ & \thead{Previous anti-commuting \\ measurements} \\ \hline
1 & $xxx$ & 0.428 & --- \\
2 & $yzx$ & 0.000 & --- \\
3 & $xzz$ & -0.285 & --- \\
4 & $xzy$ & 0.000 & $xzz$ \\
5 & $zyx$ & 0.000 & $xzz$, $xzy$ \\
6 & $xzx$ & -0.602 & $xxx$, $yzx$, $xzz$, $xzy$ \\
7 & $xxz$ & -0.602 & $xxx$, $yzx$, $xzz$, $zyx$ \\
8 & $zxx$ & -0.602 & $xxx$, $xzz$, $xzy$, $zyx$ 
\end{tabular}
\end{center}

This trace shows that the forest algorithm follows the tree algorithm's strategy of
scanning measurements from a mutually commuting subset until correlation 
complimentarity indicates switching to a different subset. However, the random
forest's thresholds for switching are more finely tuned compared to an arbitrary constant
value of $\<B\>^2 > 0.4^2$ for the tree algorithm. In the example, even though
the first measurement is slightly above the threshold of $0.4^2$, the forest algorithm
comes back to measurements that anti-commute with $xxx$ and finds three large
measurements to declare the state entangled. However, once the forest algorithm 
finds $xzx$ and $xxz$, it avoids measurements that anti-commute with them.

In some sense, the collection of forests is a set of non-parametric models capturing correlation 
complimentarity through the joint probability distribution of $\overline{x}(\rho)$. 
Then, the tree scoring function can then be viewed as a 
conditional expectation. This is more clear when rewriting Eq. (\ref{tree_score}) as:
\begin{equation}
\frac{1}{\sum_{r_i \cap R_i \neq \emptyset} p(r_i) + n(r_i)} \cdot \left[
	\sum_{r_i \cap R_i \neq \emptyset} 1 \cdot p(r_i) + 0 \cdot n(r_i)
\right]
\end{equation}
Since the expectation of an indicator is the probability of the indicated event. Thus,
the scores represent a direct estimate of the probability that an unknown
measurement would be the best at declaring entanglement in fewer steps.

\section{Conclusion}
This paper presents an example usage of machine learning in entanglement 
detection. Empirical evidence shows that the forest algorithm is an improvement, for 
two to six qubits, over the existing tree algorithm. The forest algorithm achieved this 
without sacrificing ``correctness'': every state the forest algorithm declares 
entangled is guaranteed to be entangled, even though it may take longer or 
fail to detect entanglement.

Developments in machine learning provide non-parametric methods for analysing
quantum systems. Non-parametric methods can eliminate arbitrary parameters in
analyses. The success of the forest algorithm is an example where the analysis was
improved by eliminating the threshold to move onto a different mutually commuting
subset. It is possible this approach can be applied to other problems in quantum
information processing.

{\bf Acknolwedgements}. 
The author thanks Debbie Leung for help on the concentration
of measure and general support throughout the writing of the manuscript. 
A large part of this work was completed while the author was a student at 
the University of Cambridge and the author is grateful for the supervision 
received from Stephen Brierley and Anuj Dawar.

\bibliographystyle{quantumbib.bst}
\bibliography{entanglement-v2}

\end{document}